\documentclass[a4paper]{article}
\usepackage[utf8]{inputenc}
\usepackage[]{todonotes}
\usepackage{booktabs}
\usepackage{siunitx}[=v2]
\newcommand{\ra}[1]{\renewcommand{\arraystretch}{#1}}

\usepackage{xcolor}
\usepackage{colortbl}
\usepackage[a4paper,left=1.0in, right=1.0in]{geometry}
\usepackage[sc]{mathpazo}
\usepackage{multicol}
\usepackage{multirow}
\usepackage{makecell}
\usepackage{array}
\usepackage{pdflscape}
\usepackage{hyperref}
\usepackage{subcaption}
\usepackage{lineno}

\author{
Oskar Hartbrich${}^*$, Umberto Tamponi${}^+$, Gary S. Varner${}^*$\\
${}^*$University of Hawaii at Manoa,${}^+$INFN Torino
}
\title{STOPGAP - a Time-of-Flight Extension for the Belle\,II TOP Barrel PID System as a Demonstrator for CMOS Fast Timing Sensors}
\date{2021}

\begin{document}
\maketitle


\begin{abstract}

The Belle\,II barrel region is instrumented with the Time of Propagation (TOP) particle identification system based on sixteen fused-silica bars arranged around the interaction point acting as Cherenkov radiator. Due to the mechanical design of the TOP system these quartz bars do not overlap, but leave a gap of around \SI{2}{\cm} between them. This leads to around \SI{6}{\percent} of all tracks in the nominal TOP acceptance region to escape without traversing any of the quartz bars and thus not giving any usable particle identification information from the TOP system and an additional \SI{3}{\percent} of tracks being degraded due to edge effects.
We propose a possible solution to remedy these gaps in the TOP acceptance in the form of a Supplemental TOP GAP instrumentation (STOPGAP) that covers the dead area between adjacent quartz bars with fast silicon detectors to directly measure the time-of-flight of traversing particles for particle identification purposes. Modern, fast timing silicon sensors and readouts can offer sufficient time resolution for the task at hand, so that STOPGAP modules could be built compact enough to fit into the limited space available in the area of interest between the Belle\,II central drift chamber (CDC) and the TOP system. 

In this article, we present a simulation study demonstrating the feasibility of a silicon time-of-flight system based on its reconstruction performance in $\Upsilon(4S)\rightarrow B\bar{B}$ events simulated using the Belle\,II simulation and reconstruction software. We discuss the performance requirements for possible sensor technologies and demonstrate that such a project could be realised with novel, fast monolithic CMOS sensors that are expected to reach MIP timing resolutions of down to \SI{50}{\ps}. Additionally, we discuss the use of fast timing layers at lower radii for track triggering as well as particle identification at low momenta \SI{<1}{GeV/c}.
\end{abstract}

\section{Introduction}
The physics program of super B-factories requires the best possible reconstruction of all final state particles in a given event. The differentiation between charged particles species (electrons, muons, pions, kaons, protons) is accomplished by dedicated particle identification (PID) subdetectors and further assisted by their ionization energy loss ($dE/dx$) as measured in the tracking system as well as their energy deposition in the electromagnetic calorimeter and the muon system. 
The Belle\,II barrel region is instrumented with the Time of Propagation (TOP) particle identification system, which measures the detection time of Cherenkov photons generated sixteen \SI{2625x450x20}{\mm} quartz radiator bars arranged around the interaction point (IP). The photons are trapped inside the bars by total internal reflection and transmitted to an array of pixelated photo-sensors (MCP-PMTs). The measured time difference between the particle collision and the photon arrival time is the sum of two contributions: first, the particle's time of flight from the IP to the quartz bar, which is inversely proportional to the particle velocity. Second, the photon propagation time inside the quartz bar, which is a function of the Cherenkov angle \cite{Belle2, topfee} and thus the particle velocity. The particle crossing the bar is identified comparing the time distribution of the photons detected in each photo-sensors pixel with the expected ones for different mass hypotheses.
The TOP is responsible for most of the for hadron identification capability in the barrel region, which corresponds to $\approx 80\%$ of the total Belle II acceptance. 
The mechanical design of the TOP modules leaves gaps in the azimuthal direction ($\phi$) of about \SI{2}{\cm} width in between the individual quartz bars, accounting for about \SI{6}{\percent} of missing geometric coverage in the nominal TOP acceptance. Tracks passing through the outermost edges of a TOP bar also have reduced particle identification performance, effectively widening this gap to around \SI{16}{\percent} of tracks with no or degraded TOP PID. Many physics analyses foreseen in the Belle\,II physics program require the positive identification of several final state particles, multiplying the impact of the TOP quartz gap. A particularly important case is represented by the analyses based on the full event reconstruction (FEI) with the tagging of the opposite B meson, like the semileptonic B decays with undetected neutrinos. Such decays are of particular interest for searches of phenomena beyond the standard model and, in some cases, can be uniquely explored at Belle\,II. Flavor tagging, used for CP violation measurements, also depends on the efficiency and quality of lepton and hadron identification. 

 The TOP does not only contributes to the particle identification, but plays  a central role in the global Belle II reconstruction. A most critical component for both particle identification and tracking in the silicon vertex detector is the event time $T_0$. The TOP detector is the most precise timing instrument in the Belle\,II apparatus and thus plays an important role in its determination. However, the TOP contribution to the $T_0$ measurement strongly depends on the number of tracks in the TOP acceptance. Especially with the expected increase in background rates, the impact of a precise $T_0$ determination will continue to grow in importance. Thus, not only the particle identification, but the overall quality of the Belle\,II event reconstruction would be improved by extending the TOP coverage.

A possible solution to remedy the gaps in the TOP acceptance is to install a supplemental time-of-flight detector that covers the non-instrumented areas between adjacent quartz bars. Since the available space around the installed TOP modules is quite limited, one or multiple layers of fast-timing silicon detectors would serve this scope. The Supplemental TOP Gap Instrumentation (STOPGAP) has become feasible with the advent of modern silicon sensor types with very short signal collection times and excellent time resolutions in the range of a few tens of picoseconds, which we will show to be sufficient to provide particle identification in the momentum range \SIrange{0.05}{5.0}{\GeV\per c} and path lengths of \SIrange{1}{2}{\m} at super B-factory experiments such as Belle\,II.

\begin{figure}[htbp]
	\centering
		\includegraphics[width=0.8\textwidth]{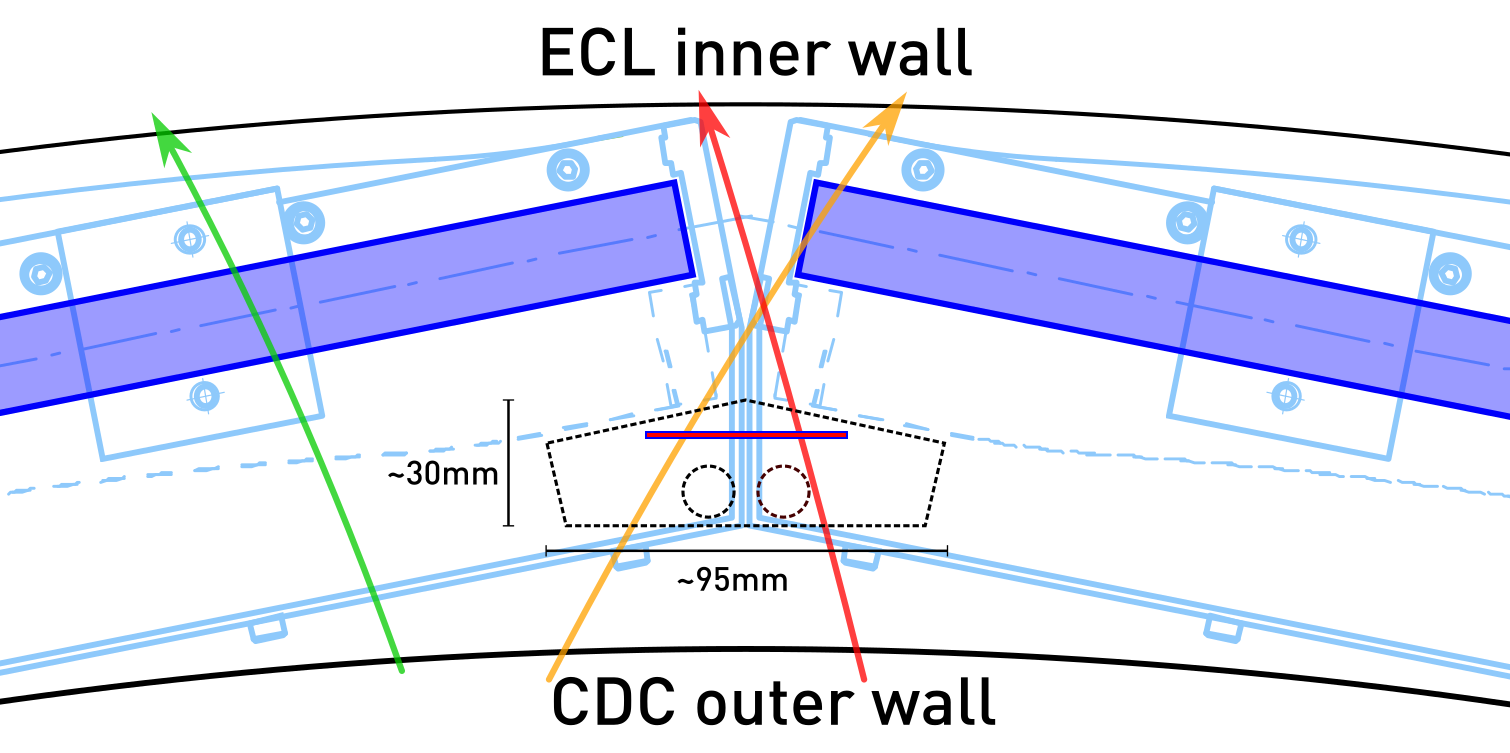}
	\caption[]{Conceptual sketch of supplemental TOP gap instrumentation module in the cross section of two adjacent TOP modules as seen from the backward side. The solid light blue lines show the outlines of the readout side of the TOP quartz bar box. The dashed light blue lines show the outline of the rest of the TOP quartz bar box. The TOP quartz cross section is shown in dark blue. The dashed black lines roughly show the dimensions of a possible STOPGAP module, with the two dashed circles indicating possible cooling lines or cabling channels and a layer of silicon sensors shown in pink.}
	\label{fig:schematic_view}
\end{figure}

There is around \SI{45}{\mm} of free space between the the outer shell of the Belle\,II Central Drift Chamber (CDC) and the inner wall of the TOP module enclosures. This is sufficient to fit two layers of silicon sensors \SI{40}{\mm} wide, and the additional required services, to cover the gap between the TOP quartz bars along their whole length. Fig.\ref{fig:schematic_view} shows a sketch of the geometry around the gap between two TOP module enclosures and a possible STOPGAP module geometry.

\section{MC Performance Study}
\begin{figure}[htbp]
	\centering
    \includegraphics[width=0.7\textwidth]{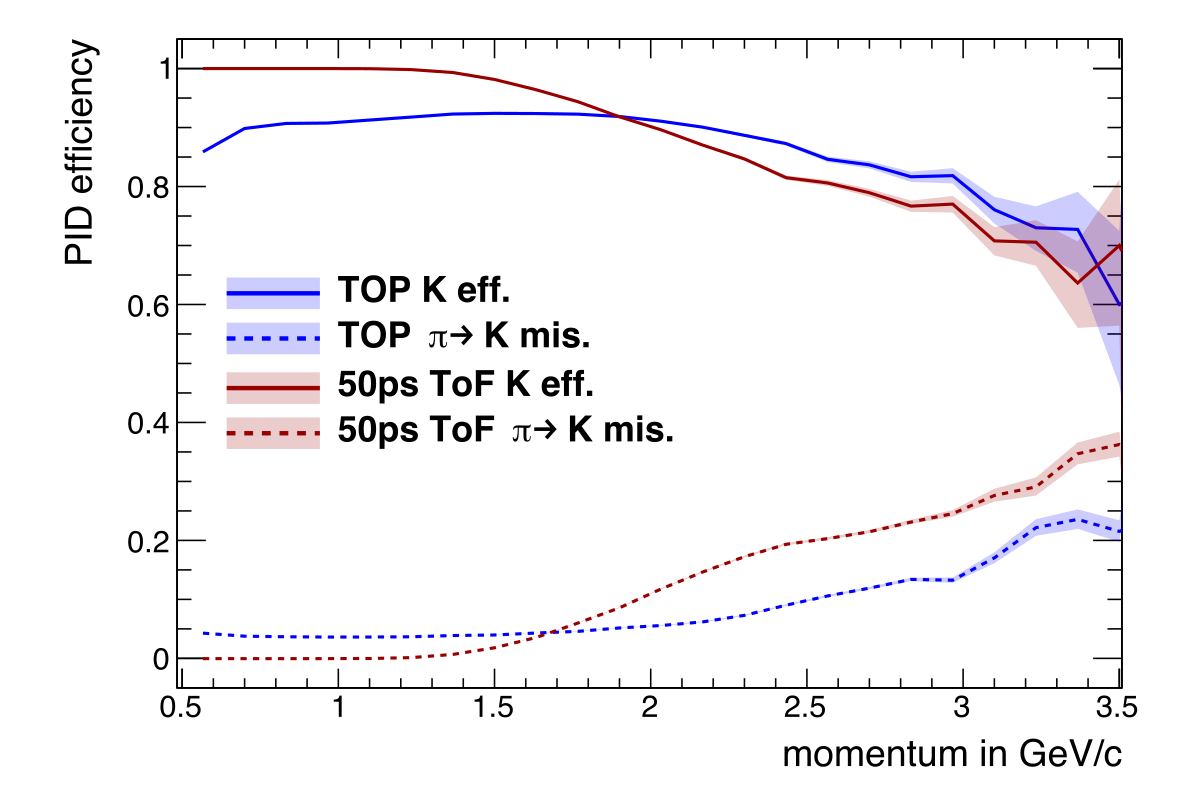}
	
	\caption[]{Selection and mis-identification probabilities at $LL_{K}-LL{\pi}>0$ of a simulated STOPGAP PID system for charged kaons as a function of particle momentum, shown for a simulated sensor with \SI{50}{ps} MIP timing resolution with additional timing contributions as explained in the text. Solid curves are the kaon selection efficiencies, while the dashed curves show the mis-identification probability for $\pi\rightarrow K$. Shaded areas indicate the statistical uncertainties of the data points.}
	\label{fig:tof_perf}
\end{figure}

We performed Monte Carlo (MC) study of the STOPGAP performance using simulated $\Upsilon(4S) \to B\bar{B}$. The existing structures of Belle II detectors are simulated using Geant4, and the charged particles' trajectories are reconstructed using the  publicly available Belle II analysis software framework \cite{basf2, basf2_github}. 
The STOPGAP response is calculated using a parametric MC approach using the extrapolated impact point of the track of the inner TOP surface, which approximately corresponds to the planned radial position of the STOPGAP modules. 
To simulate the STOPGAP response several contributions to the total time resolution have been considered, both reducible and irreducible: sensor readout resolutions from \SIrange{20}{100}{\ps}, global clock distribution jitter (\SI{10}{\ps}), the SuperKEKB bunch overlap time (\SI{15}{\ps}), and track length uncertainty, parameterized assuming a \SI{5}{\mm} resolution on the impact point of the track in the Z direction. The particle identification is then performed by evaluating a likelihood constructed from the these time resolution components, all assumed to be Gaussian. The resulting selection and mis-identification fractions for charged pions and kaons are shown for a sensor resolution of \SI{50}{ps} in Fig.\ref{fig:tof_perf}. Detailed plots for both pion an kaon (mis-) identifications are attached Figs.\ref{fig:tof_perf_kaon},\ref{fig:tof_perf_pion} in the appendix. A STOPGAP system with a combined time resolution for sensor and readout of \SI{50}{\ps} would perform significantly better than the TOP system in the momentum range $p<\SI{2}{\GeV}$. A \SI{30}{\ps} system would outperform TOP in the whole momentum range.

\begin{figure}[htbp]
	\centering
\includegraphics[width=0.7\textwidth]{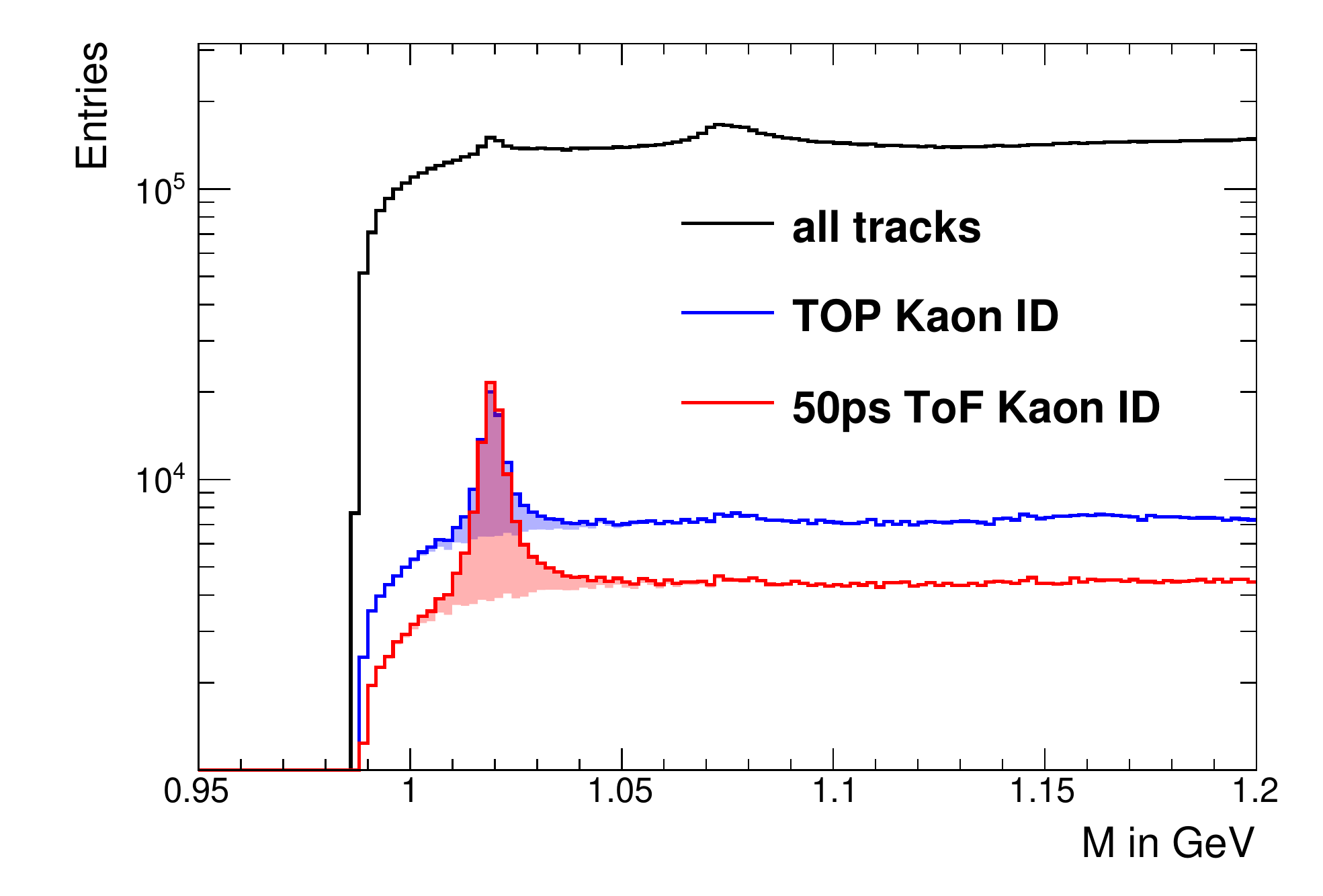}
	\caption[]{Invariant mass distribution from charged kaons selected at $LL_{K}t>LL{\pi}$ for both tracks, with both tracks in the TOP acceptance. The black line shows the unselected spectrum. The shaded areas indicates the fraction corresponding to true $\phi$ decays. The TOP system achieves a S/N of 0.37 ($\mu\pm2\sigma$), while the STOPGAP reconstruction achieves a S/N of 0.79 in the same invariant mass window.}
	\label{fig:phi_KK}
\end{figure}

Fig.\ref{fig:phi_KK} shows the reconstructed invariant mass of $\phi\rightarrow K^+ K^-$ decays selected from the simulated dataset. A \SI{50}{\ps} STOPGAP system achieves more than twice better signal-to-noise ratio near the $\phi$ mass peak. The improved efficiency is entirely defined by the difference in kaon efficiency, which is only \SI{90}{\percent} for TOP, but driven by the MIP efficiency in STOPGAP.

The next step will be the implementation of a realistic STOPGAP geometry into the Belle\,II simulation, which will allow us to study the combined performance of TOP+STOPGAP as well as its impact on specific physics channels of interest. This would also allow realistic studies of the resilience of a time of flight system vs. beam backgrounds. The expected improvement in the combined particle identification has to be understood to optimize the number of  sensor layers and their ideal dimensions and arrangement.

\subsection{Backgrounds}
We identified two main sources of background affecting STOPGAP: backsplash from showers in the electromagnetic calorimeter located few centimeters behind the TOP, and beam-induced backgrounds. We expect the former to be easily identifiable in the majority of cases, since backsplash hits are necessarily later in time that the initial particle hit and in addition should deposit significant energy in the corresponding ECL crystal. 
The beam background are discussed in \autoref{sec:hitrates}, where in lieu of a true GEANT4-based simulation study,  we use the occupancy in the vertex detector as a proxy for the background level and perform some extrapolations for a few conservative sensor parameters to show that the expected impact is minimal.

\section{Timing Layers at Lower Radii}
Several of the currently proposed and discussed upgrades of the Belle\,II vertexing system plan to increase the inner radius of the current drift chamber replacing its inner layers with silicon sensors. The thin silicon sensors proposed for the vertexing upgrade are unlikely to provide enough $dE/dx$ discrimination for low momentum particle as the CDC does currently. Additionally, the CDC currently provides track triggering for transverse particle momenta $p_T$ down to around \SI{100}{\MeV}, which a full silicon inner tracking system might not be able  to provide. We thus explore here the potential of recovering the PID and track triggering capabilities for low $p_T$ particles by analytical modelling of dedicated timing layers at \SI{250}{\mm} and/or \SI{450}{\mm} radius.

For pion/kaon separation purposes, the region of $p_T<\SI{500}{\MeV}$ is most relevant, as higher $p_T$ will reach the dedicated PID subdetectors TOP, ARICH and also generally provide a $dE/dx$ measurement from the remaining part of the CDC or a potential TPC replacement. We calculate the time of flight from the IP until the first crossing of the timing layer radius as a function of particle momentum and mass, including the path length of the circular segment that describes its trajectory. In order to keep the calculation simple and conservative, we assume $\left|p\right| = p_T$ here. 

\begin{figure}[htbp]
	\centering
    \includegraphics[width=0.7\textwidth]{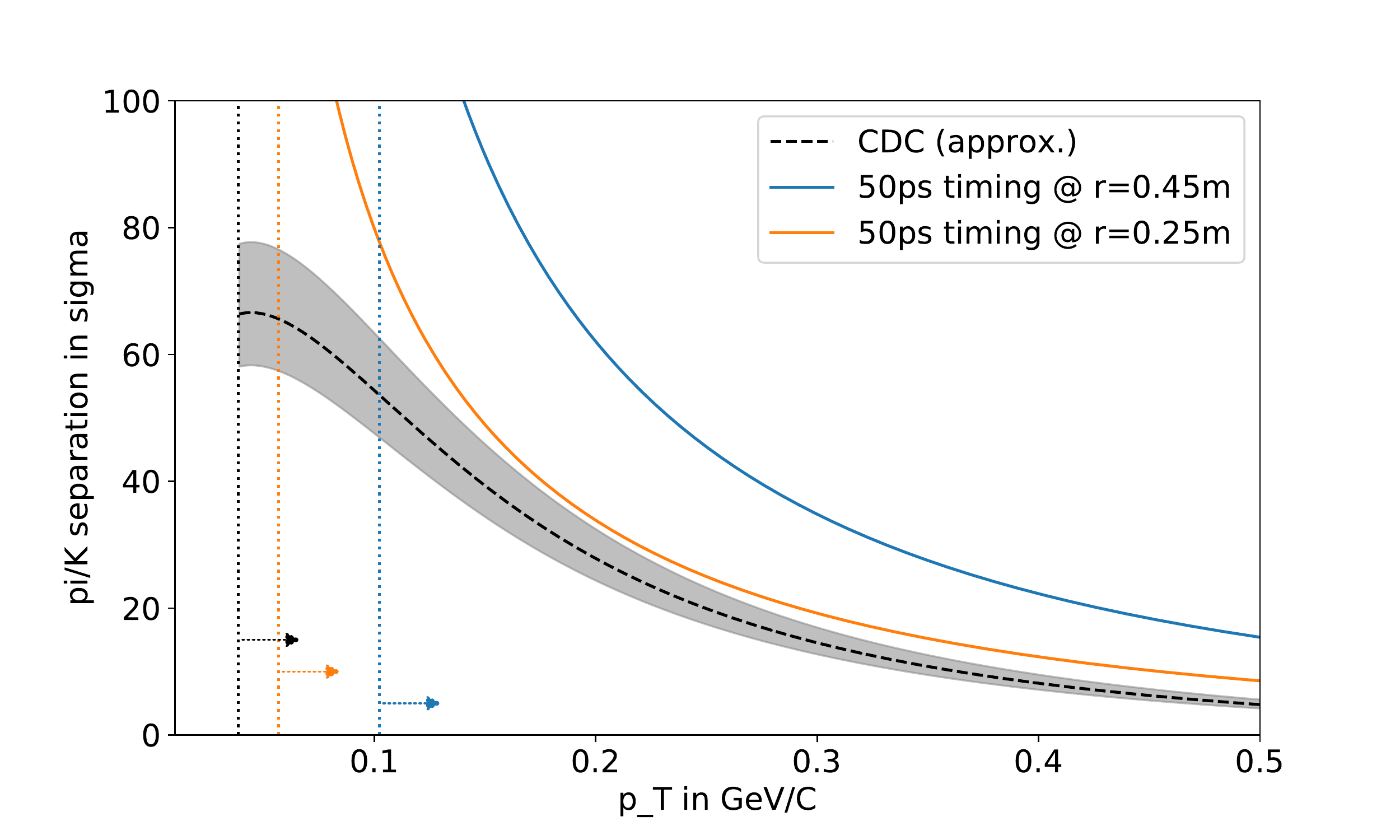}
	
	\caption[]{Pion/kaon separation performance of a MIP timing layer compared to an estimated $dE/dx$ PID at different radii. The arrows indicate the lower $p_T$ cutoff for each system due to its radius.}
	\label{fig:tof_lowpt}
\end{figure}

The resulting time of flight differences  between charged kaons and pions are then divided by the assumed time resolution of a timing layer, resulting in a measure or particle identification in standard deviations of separation. These numbers are then compared to an estimate of the $dE/dx$ particle separation of the CDC as shown in Fig.\ref{fig:tof_lowpt}. This study indicates that a \SI{50}{\ps} MIP timing resolution layer at \SI{250}{\mm} radius yields consistently better pion/kaon separation than the CDC, while a \SI{450}{\mm} radius timing layer leads to at least double the separation power between pions and kaons over the CDC in the momentum region inspected here.

To estimate the general feasibility of timing layers to provide track triggers for Belle\,II, simplified trigger model is combined with extrapolations from the beam background estimates described in \autoref{tab:backgrounds} below. The individual layer geometry is assumed as a cylinder length of three times its radius, leading to a covered area of \SI{1.2}{\meter\squared} at \SI{250}{mm} radius or \SI{3.8}{\meter\squared} at \SI{450}{\mm} radius.

For each examined timing layer radius, the expected background hit rates at full luminosity are extrapolated from all given radii and their ensemble mean and min-max envelope used for further calculations. An additional simplifying assumption made here is that background hits are spatially uncorrelated.

A timing system at the trigger level would likely not work at the full granularity of the used sensor technology, but instead operate on larger areas for coincidence. We thus conservatively assume here that the effective "trigger region" of coincidence of such a system is \SI{10}{\cm\squared}. Such regions would be formed by OR'ing the outputs of all individual sensor channels in that region. For the purposes of this study a "track" consists of $N$ stacked (or geometrically associated) trigger regions firing coincident in time within a conservative window of \SI{1}{\ns} (corresponding to a \SI{\pm5}{\sigma} range of an assumed \SI{100}{\ps} MIP timing resolution at the trigger level). Individual tracks are considered coincident when $M$ track occur within a time window of \SI{5}{\ns}, which roughly covers the time of flight differences between electrons and protons at the relevant radii and momenta here.

Based on these assumptions, the trigger rates from pure beam backgrounds can be calculated. The event $T_0$ reconstruction at the trigger level is not expected to reach the resolution of the timing layers discussed here, so a single timing layer can only form a coincidence with itself. This is prohibitive in expected noise rates, so we only investigate cases of different radial configurations of two timing layers forming a coincidence.

\begin{figure}[htbp]
	\centering
    \includegraphics[width=0.7\textwidth]{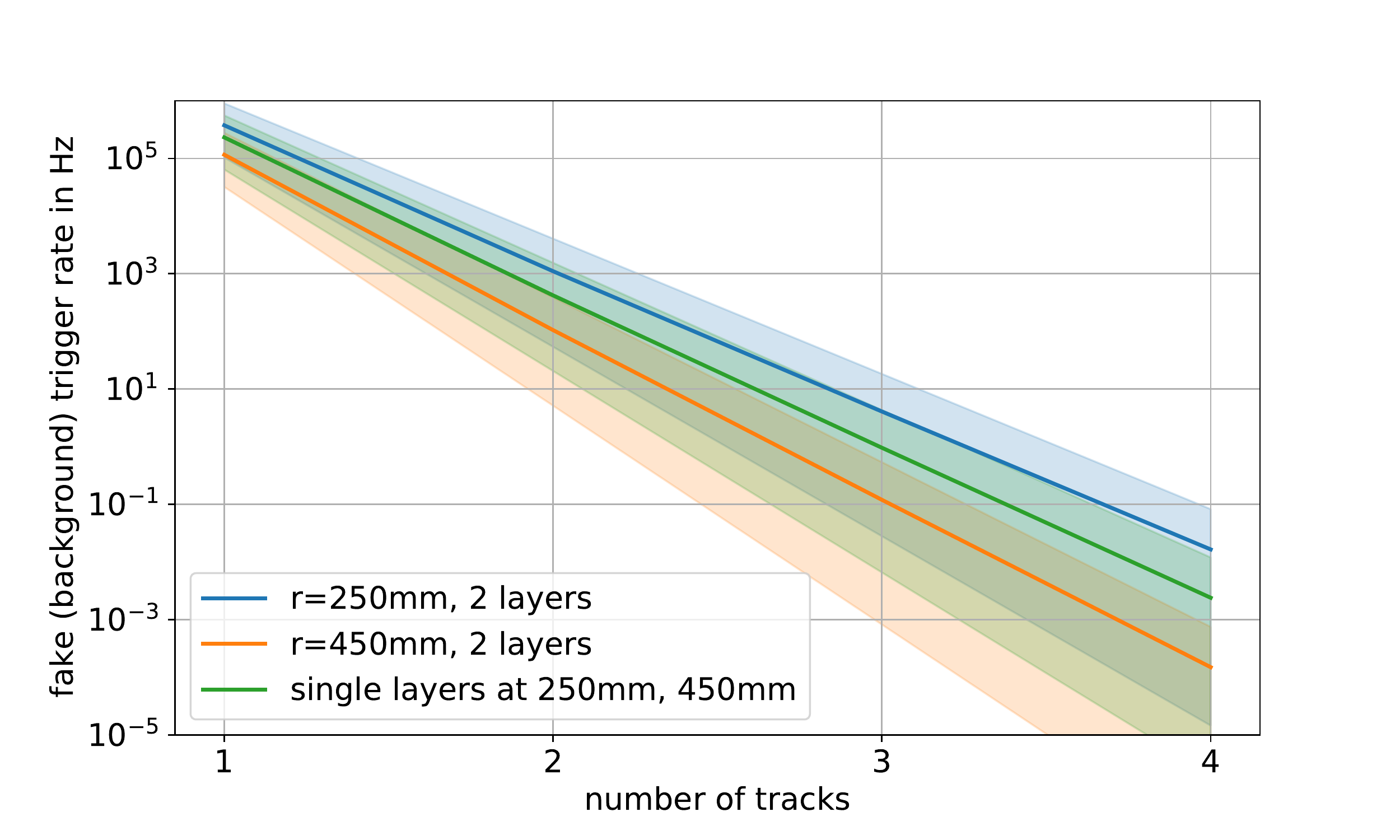}
	
	\caption[]{Estimated background rate of double layers of fast timing coincidence track trigger at \SI{250}{\mm} and \SI{450}{\mm} radii as well as for two single track trigger timing layers at \SI{250}{\mm} and \SI{450}{\mm} radius.}
	\label{fig:timing_trig_twolayer}
\end{figure}

Fig.\ref{fig:timing_trig_twolayer} shows the results of the model described above for the beam backgrounds expected at full design luminosity of SuperKEKB. Depending on the radius, each additional track required for a trigger decreases the background trigger rate by 2-4 orders of magnitude. Increasing the dual timing layer radius from \SI{250}{\mm} to \SI{450}{\mm} decreases the background trigger rates by at least one to two orders of magnitude. 

Notably, a configuration with two single timing layers at \SI{250}{\mm} and \SI{450} radius, respectively, is estimated to yield background trigger rates right in between the double-layer options. However, the base assumption of spatially uncorrelated background hits is much more likely to be accurate in this configuration. Additionally a distanced two-layer setup would enable a coarse estimation of the Z-origin of a triggered track, enabling a selection of tracks from the IP, as well as a coarse estimation of the track momentum (or at least its charge) on the trigger level at the cost of increased complexity in the required trigger reconstruction logic.

The uncertainties from the beam background extrapolations generally span at least an order of magnitude, and the systematic uncertainties from the assumptions described above are not included in the figure at all. The given numbers nevertheless show the fundamental feasibility of a timing track trigger system, at least for triggering events with two or more tracks in its acceptance. 

\section{Requirements}
We now review the requirements for a STOPGAP upgrade to the Belle\,II TOP detector system or even its full replacement. The technological requirements for timing layers could be extracted in the same general way, but are not discussed in further detail here.

Depending on the details of the chosen STOPGAP geometry, \SIrange{1}{3}{\m\squared} of active area will be needed to fill the gaps in the TOP acceptance with a single sensor layer. In case a double-layer design is necessary, that area would double accordingly. For reference, a full replacement of the TOP system would require around \SI{18}{\m\squared} of timing sensitive sensor.

As was described earlier, the particle identification performance of a time-of-flight system is fundamentally limited by the single MIP detection efficiency of the used sensor technology. A MIP efficiency as close as possible to \SI{100}{\percent} is thus a strong requirement for such a system.

The allowable material budget for a STOPGAP module should generally not exceed the material budget of a full TOP module consisting of several \si{\mm} of aluminium and around \SI{2}{\cm} of quartz (about \SI{0.2}{X_0}). Since there will be some effective overlap between the two structures, minimising the STOPGAP material budget will be helpful in reducing potential impacts on the photon reconstruction from conversions in the material in front of the ECL crystals.

In the following, the expected background hit rates and radiation dosages are discussed. This discussion is based on extrapolations of the expected beam background rates and radiation dosages - without additional safety factors - at full SuperKEKB luminosities at the radii of the currently installed Belle\,II vertex detector \cite{baudot}. The given numbers are extrapolated to the radius of the TOP system and potential radii of timing trigger layers. The original values as well as our radial extrapolations are given in \autoref{tab:backgrounds}.

\begin{table}
    \centering
    \caption{Estimated hit rates and radiation dosages at full SuperKEKB luminosity of \SI{8E35}{\per\cm\squared\per\second} for the existing Belle\,II VXD layer radii \cite{baudot}. The given numbers for a STOPGAP system or timing trigger layers at \SIrange{250}{450}{\mm} are extrapolated from the given ensemble of VXD estimates. The given range for each radius is the minimum and maximum value obtained from the ensemble.}
    \ra{1.1}
    \begin{tabular}{lrrrr}
    \toprule
    Layer & Radius & Hit Rate & NIEL & Total Ionising Dose\\
     & in \si{\mm} & in \si{\Hz\per\cm\squared} & in \si{n_{eq} \per \cm \squared} & in \si{Rad} \\
    \midrule
    1 (PXD) & \num{14} & \num{22.6E6} & \num{10E12} & \num{2E6} \\
    2 (PXD) & \num{22} & \num{11.3E6} & \num{5E12} & \num{600E3} \\
    3 (PXD) & \num{39} & \num{1.41E6} & \num{0.2E12} & \num{100E3} \\
    4 (SVD) & \num{80} & \num{290E3} & \num{0.1E12} & \num{20E3} \\
    5 (SVD) & \num{115} & \num{220E3} & \num{0.1E12} & \num{10E3} \\
    6 (SVD) & \num{135} & \num{150E3} & \num{0.1E12} & \num{10E3} \\
    \midrule
    
    STOPGAP & \num{1167} & \makecell[r]{2416 \\ (\numrange{1362}{4015}) } & \makecell[r]{\num{1.05E9} \\ (\numrange{0.22E9}{1.78E9}) } & \makecell[r]{158 \\ (\numrange{94}{288}) } \\
    
    Trigger 1 & \num{450} & \makecell[r]{\num{16.2E3} \\ (\numrange{9.2E3}{27E3}) } & \makecell[r]{\num{7.08E9} \\ (\numrange{1.50E9}{12.0E9}) } & \makecell[r]{1062 \\ (\numrange{632}{1936}) } \\

    Trigger 2 & \num{250} & \makecell[r]{\num{52.6E3} \\ (\numrange{29.7E3}{87.5E3}) } & \makecell[r]{\num{23.0E9} \\ (\numrange{4.87E9}{38.7E9}) } & \makecell[r]{3442 \\ (\numrange{2048}{6272}) } \\

    \bottomrule
    \end{tabular}
    \label{tab:backgrounds}
\end{table}

\subsection{Estimation of STOPGAP hit rates} \label{sec:hitrates}
Based on \autoref{tab:backgrounds}, we calculate a STOPGAP background rate of \SI{2.4+-0.9}{\kHz\per\cm\squared} at the SuperKEKB design luminosity of \SI{8E35}{\per\cm\squared\per\second}. In order to be conservative, we assume an additional safety factor of 5, and an extra luminosity factor of 5 for an assumed maximum background hit rate of \SI{60}{\kHz\per\cm\squared}. Further assuming a conservative readout open time window of \SI{100}{\ns} (to broadly cover an assumed trigger time jitter of \SI{10}{\ns} and a maximum time of flight of individual particles of \SI{20}{\ns}), we arrive at a background occupancy of $\SI{0.60+-0.23}{\percent\per\cm\squared}$. Finding the hit on the STOPGAP sensor that corresponds to a given track needs to consider around \SI{1}{\cm\squared} of sensor area due to the limited Z-pointing resolution of the CDC tracking. We thus arrive at a very conservatively estimated chance of \SI{0.6}{\percent} that there is a background hit in the same spatial and temporal region of interest on the STOPGAP surface for a given charged particle track. 

\subsection{Radiation Hardness of STOPGAP}
As given in \autoref{tab:backgrounds}, the expected non-ionizing lattice displacement (NIEL) at the STOPGAP radius is on the order of \SI{1E9}{n_{eq}\per\cm\squared}. This value is compatible with the NIEL estimates obtained as part of the original TOP irradiation expectation studies. Similarly, the total ionizing radiation dose at the STOPGAP radius is estimated to be \SI{0.16+-0.07}{KRad}.
These values are several orders of magnitude lower than those expected and observed in the (HL-)LHC environments. Radiation hardness is thus not expected to of significant importance in the selection of a suitable technology option.

\subsection{Granularity}
Based on the occupancy estimates and the tracking extrapolation resolution presented above, a sensor granularity of around \SI{1}{\cm^2} would be sufficient. The time resolution contribution from an uncertainty in the trajectory path length due to the Z-pointing resolution term is generally small. With sufficient granularity in Z-direction, the inclusion of STOPGAP hits into the track fit would further eliminate that uncertainty. \SI{1}{\mm} of STOPGAP sensor granularity in Z corresponds to around a \SI{1}{\ps} additional time resolution contribution from track path length uncertainty.

\subsection{Summary}
Especially in comparison to the detector upgrades under discussion and construction for the HL-LHC, the requirements on hit rates, radiation hardness and granularity are entirely negligible for a STOPGAP application. The driving factors of its performance are the single MIP timing resolution and a MIP efficiency of \SI{>99}{\percent}. As discussed above, STOPGAP modules are expected to yield useful particle identification capabilities for time resolutions \SI{<100}{ps}.

For timing layers at lower radii, the requirements on hit rate and radiation hardness are stronger than for the STOPGAP case, but still tame compared to HL-LHC developments. As part of the inner tracking detector, finer granularity and a much reduced material budget would be required compared to STOPGAP.

\section{Existing Sensor Options and Readout Technologies}
There are several options for sensors and readout technologies. We briefly review some of the existing technologies planned used for the HL-LHC upgrades and under consideration for e.g. the EIC detector concepts.

\paragraph{Low Gain Avalanche Diodes}(LGAD) are planned for the end cap timing layers of the CMS and ATLAS high luminosity upgrades \cite{atlastdr, cmstdr}. Due to their doping profiles, LGAD sensors are fabricated in specialised non-standard processes and thus rather expensive. They generally require hybridized readout chips bump bonded to them, further increasing their cost and difficulty in assembly and handling. Current LGAD designs have a large dead zone between neighbouring readout pads and hence require double layers for greater than \SI{99}{\percent} MIP efficiency. Recently first working prototypes of AC coupled LGADs have shown to remediate the efficiency gaps between pixels. Integrated readouts for LGAD sensors are under development e.g. for the ATLAS upgrade by OMEGA (ALTIROC \cite{altiroc}) and for the CMS upgrade by FNAL (ETROC \cite{cmstdr, etroc_tdc}). The target time resolution for LGAD layers is \SI{30}{ps} for a double layer. Under ideal conditions, unirradiated single LGAD layers have demonstrated down to around \SI{20}{ps} timing resolution for single MIPs.

\paragraph{LYSO+SiPM} systems were originally considered for time-of-flight positron emission tomography applications \cite{tofpet}, but are now also used for in HEP detectors, specifically the the CMS barrel timing layer (BTL) \cite{cmstdr}. The CMS BTL uses small staves of scintillating LYSO:Ce crystals around \SI{5x5x50}{\mm} read out by one silicon photomultiplier (SiPM) on each of the smaller faces. Timing resolutions of down to \SI{30}{ps} for single MIPs have been demonstrated. This approach is fundamentally limited in the possible granularity by the minimum feasible size of the LYSO crystals to around \SI{1}{\cm\squared}. The necessary material budget is limited by the required thickness of the LYSO crystal of a few \si{mm} to yield enough scintillation photons for a timing measurement of the required precision. Since SiPMs generally have large internal amplification, the noise requirements on the readout electronics are less demanding, enabling high readout channel densities per chip (see e.g. CMS TOFHiR \cite{tofhir}).

In summary, all existing fast timing sensor technologies are potentially feasible for a STOPGAP system. As a matter of fact the LYSO+SiPM CMS BTL modules would almost fit into the available space between TOP modules the way they have been designed and constructed for the CMS upgrade. However, neither the achievable granularity nor the material budget of (at the very least) a few \si{\mm} of LYSO crystals is suitable for an application at lower radii. LGADs are a potential option for both STOPGAP and track timing layers, with drawbacks in granularity, MIP efficiency and cost per area. AC-LGADs can potentially solve the first two points, but they do not address the cost issue and have also not been demonstrated in full-scale systems yet.

\section{Timing with fast MAPS}
As discussed in great details in ref.\cite{riegler}, the fundamental limitation in improving the timing resolution of silicon sensors is the achievable signal-to-noise ratio (SNR) of its output signal. LGADs address this via amplification of the ionisation charge signal in a thin avalanche amplification layer. A feasible alternative is to improve the SNR by means of a lower noise and thus higher powered electronic amplification of the pixel output signal. This has been demonstrated in principle by the NA62 Gigatracker \cite{gigatracker}, which achieves \SI{130}{ps} time resolution on single MIP hits per layer with a hybridized fast timing frontend bonded to \SI{200}{\um} thick silicon sensors without internal avalanche amplification.

\begin{figure}[htbp]
	\centering
		\includegraphics[width=0.7\textwidth]{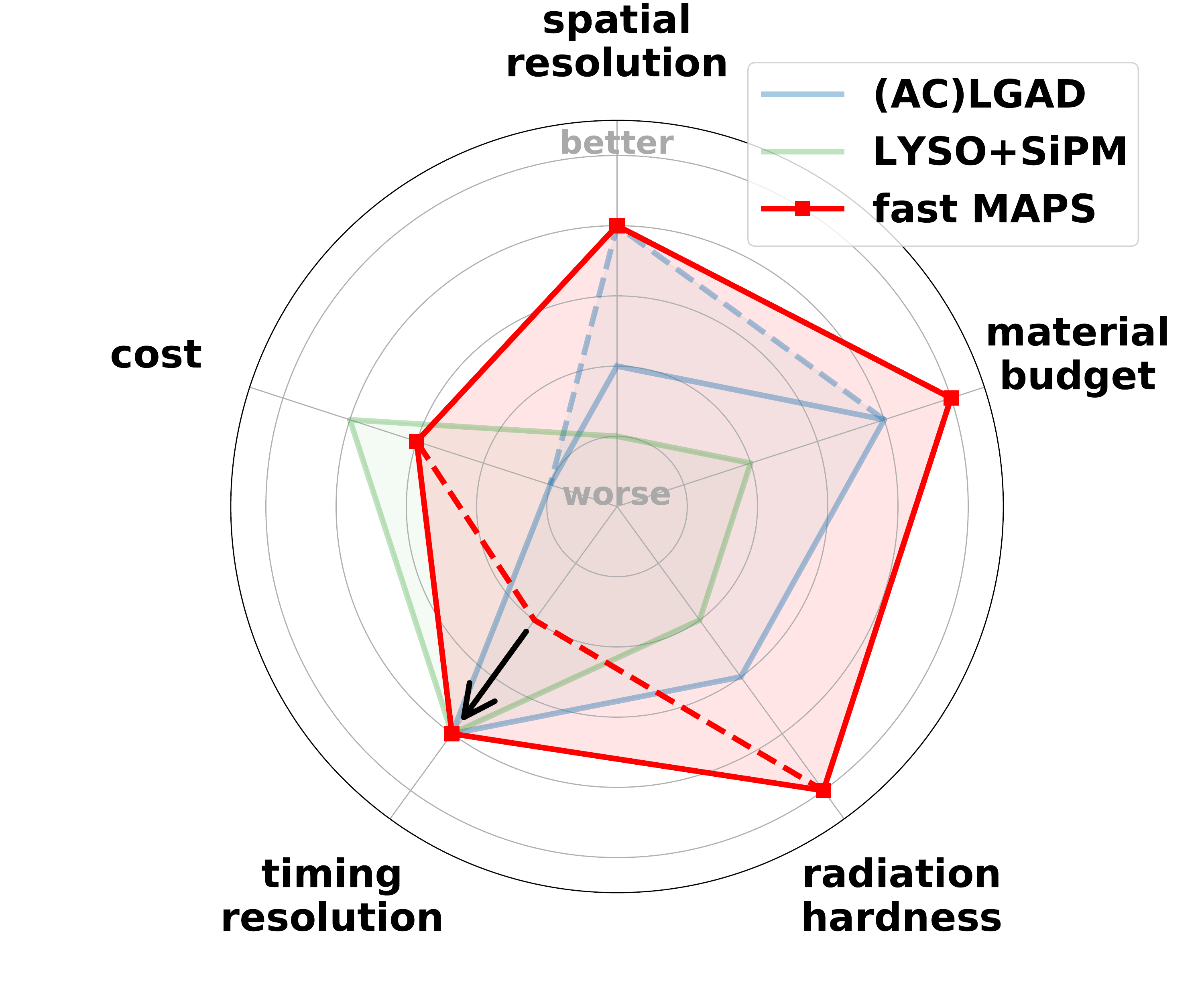}
	\caption[]{Comparison of existing fast MIP timing technologies and the potential for fast MAPS. The granularity of a LYSO+SiPM is limited in practice and the material budget is relatively high compared to silicon sensors. LGADs have shown good timing resolutions, but are expensive and somewhat limited in granularity (AC-LGAD might solve the granularity issue in the near future). Current MAPS are cost effective, highly granular and radiation hard, but their time resolution falls short by orders of magnitude compared to LYSO+SiPM and LGADs.}
	\label{fig:powerbox}
\end{figure}

The next step towards monolithic fast timing sensors is to integrate high-powered low noise preamplifiers and discriminators directly into the sensor. This is ideally produced in a CMOS process, leveraging the cost benefits and scalability of industrial CMOS fabrication. Such a chip would fully integrate the readout electronics and sparsification logic into the sensor wafer to also greatly simplify the required complexity of the electronics downstram of the sensor layer: An intelligent readout logic ideally only requires power and a reference clock as inputs, and delivers lists of hits and their timings via a fast serial output line. This would be realised in a single, thin layer of silicon with a minimum of material budget. \autoref{fig:powerbox} shows a comparison of existing fast timing technologies with the current state of MAPS and the potential for fast MAPS.

At least two groups are actively pursuing first steps towards MAPS with single MIP time resolutions of \SI{<100}{ps}. The University of Geneva is working on small pixels produced in \SI{130}{\nm}  IHP SG13G2SiGe BiCMOS \cite{geneva1, geneva2}. They have shown to achieve single MIP time resolutions down to \SI{60}{ps} in laboratory conditions for small hexagonal pixels of \SI{65}{\um} side length. In the current prototype, the power consumption necessary to achieve this time resolution is outright prohibitive at around \SI{50}{\kilo\watt\per\meter\squared}.

\begin{figure}[ht]
\begin{subfigure}[t]{0.45\textwidth}
  \centering
  \includegraphics[width=.8\linewidth]{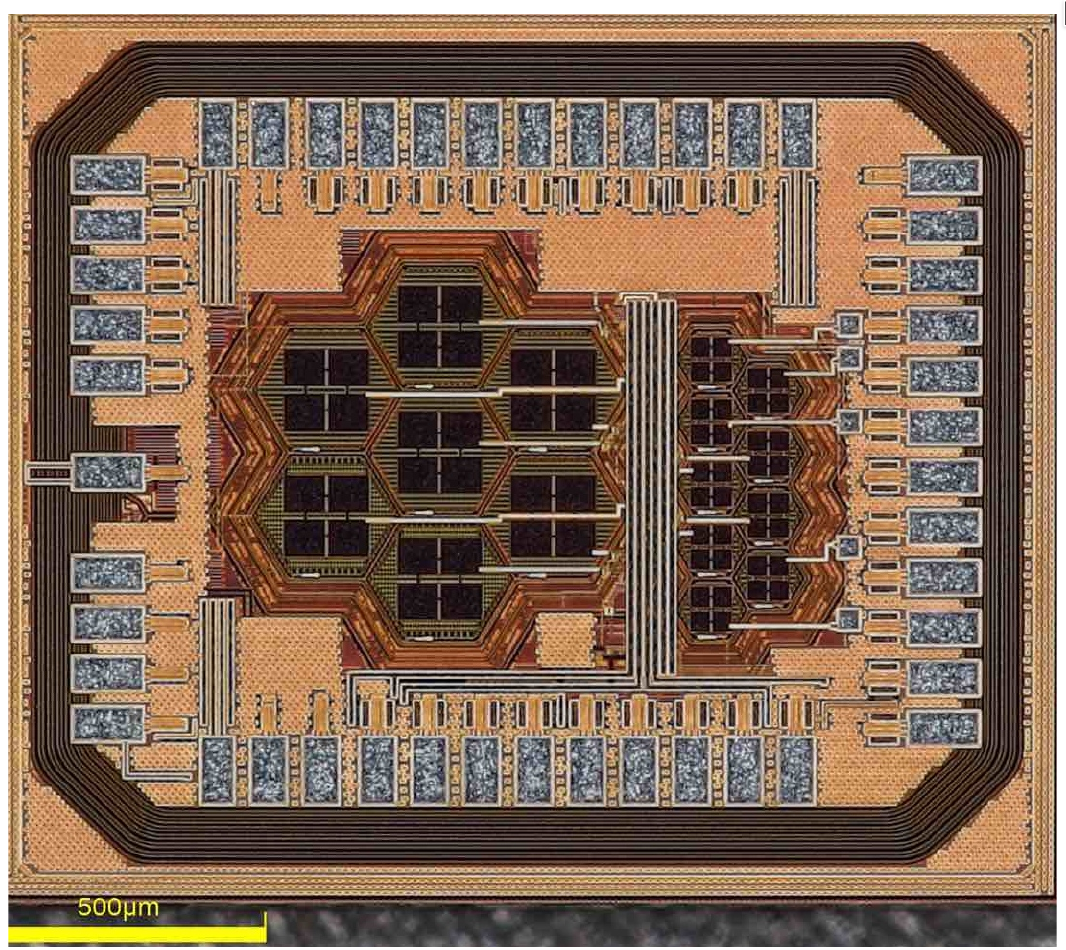}  
  \caption{Die photograph of the University of Geneva fast monolithic pixel prototype based on SiGe HBT amplifiers. The hexagonal structures are individual test pixels of \SI{130}{\um} side length (left) and \SI{65}{\um} side length (right). \cite{geneva1, geneva2}}
  \label{fig:geneva_die}
\end{subfigure}
\hfill
\begin{subfigure}[t]{.45\textwidth}
  \centering
  \includegraphics[width=.8\linewidth]{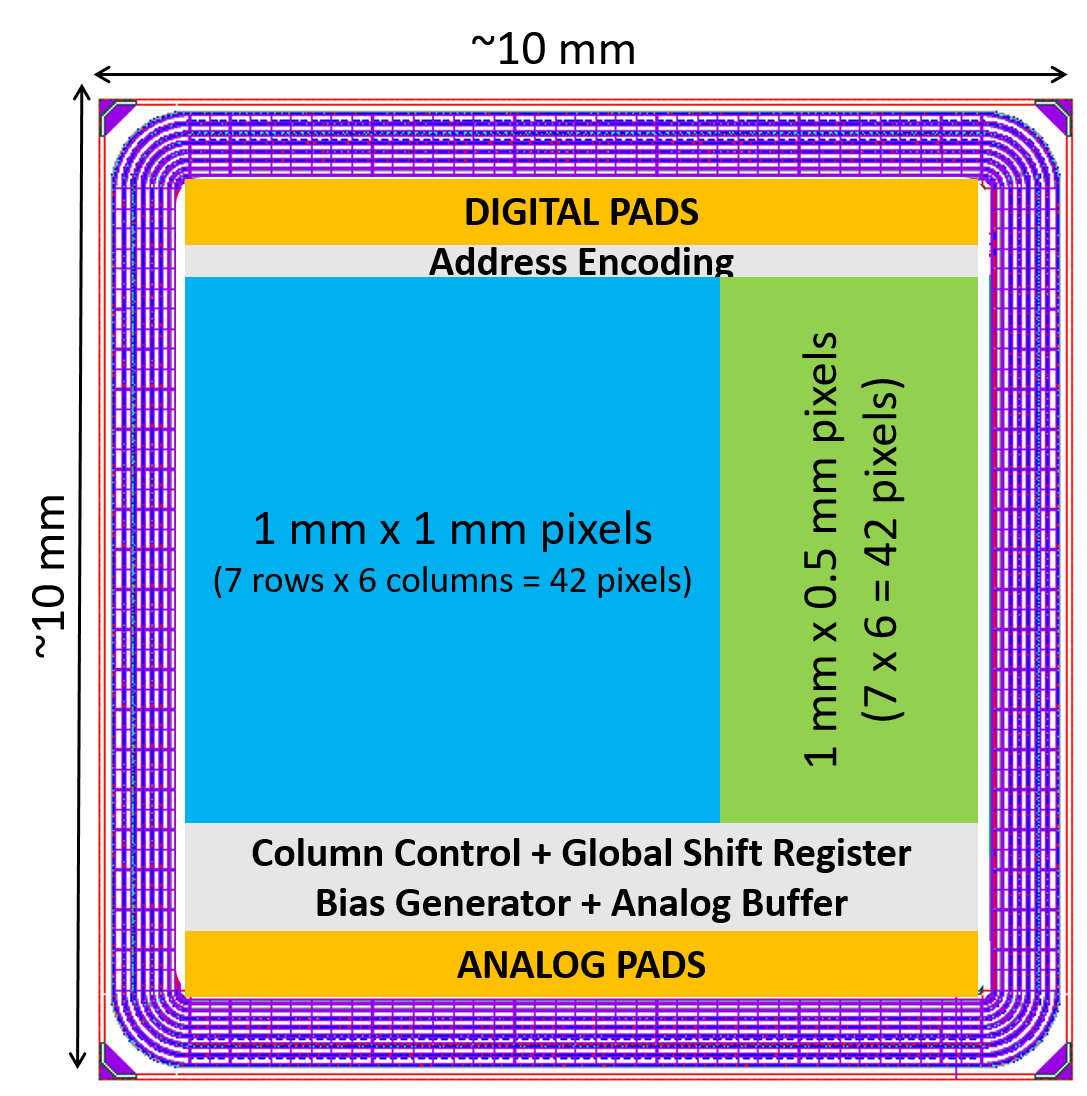}  
  \caption{Schematic of the CACTUS HVCMOS prototype chip consisting of arrays of \SI{1x1}{\mm} and \SI{1x0.5}{\mm} test pixels with additional digital logic and contact pads on the top and bottom. \cite{cactus}}
  \label{fig:cactus_diagram}
\end{subfigure}
\caption{Early fast CMOS sensor prototypes.}
\label{fig:fmaps_prototypes}
\end{figure}

The IRFU Saclay group has produced a prototype of their CACTUS concept in a \SI{150}{\nm} LFoundry HVCMOS process \cite{cactus}. The sensor is based on their experience and layout of the ATLAS monopix family, modified for the best possible time resolution. Their test structure includes \SI{1x1}{\mm} and \SI{1x0.5}{\mm} pads. Their achieved single MIP time resolutions on the first available iteration are in the order of hundreds of \si{ps}. The reason for this worse than expected performance has been understood as mis-estimated parasitic capacitances of the metal layer power rails running across the pixel. A future iteration of the same concept is expected to perform better. \autoref{fig:fmaps_prototypes} shows a picture of the Geneva prototype chip as well as a diagram of the CACTUS test structures.

These projects show that it is indeed possible to achieve sub-\SI{100}{ps} timing with monolithic CMOS detectors, even though none of the projects has produced a sensor prototype that is currently feasible for practical use.
The relatively tame requirements of a STOPGAP sensor on hit rates, radiation hardness and (in principle) readout granularity, at least compared to the requirements for HL-LHC applications, opens up STOPGAP as an ideal test bed for emerging fast timing CMOS MAPS technologies.

\begin{figure}[ht]
\begin{subfigure}[t]{0.48\textwidth}
  \centering
  \includegraphics[width=1.0\linewidth, trim=1cm 1cm 1cm 0.5cm]{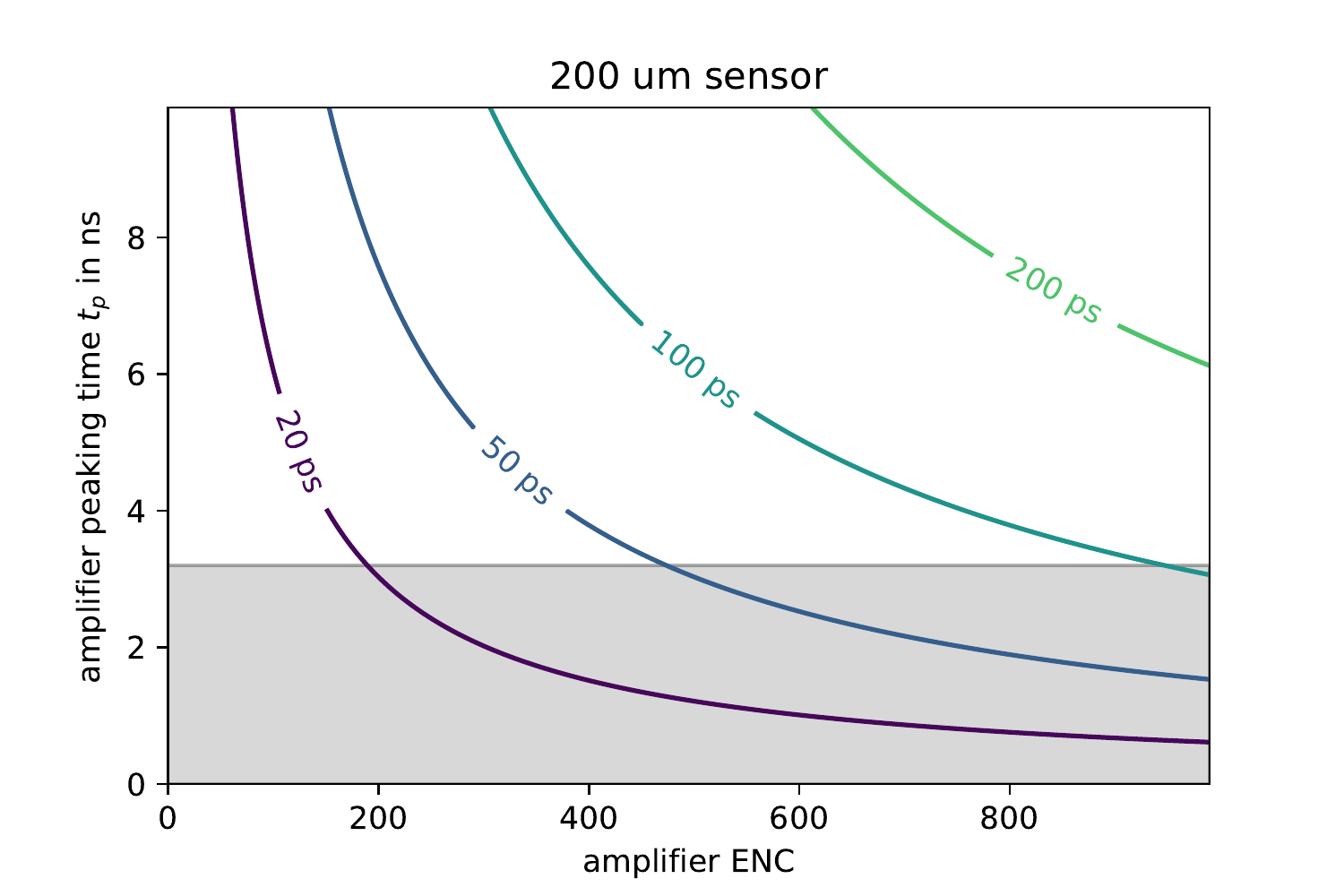}  
  \label{fig:preamp_enc_200}
\end{subfigure}
\hfill
\begin{subfigure}[t]{.48\textwidth}
  \centering
  \includegraphics[width=1.0\linewidth, trim=1cm 1cm 1cm 0.5cm]{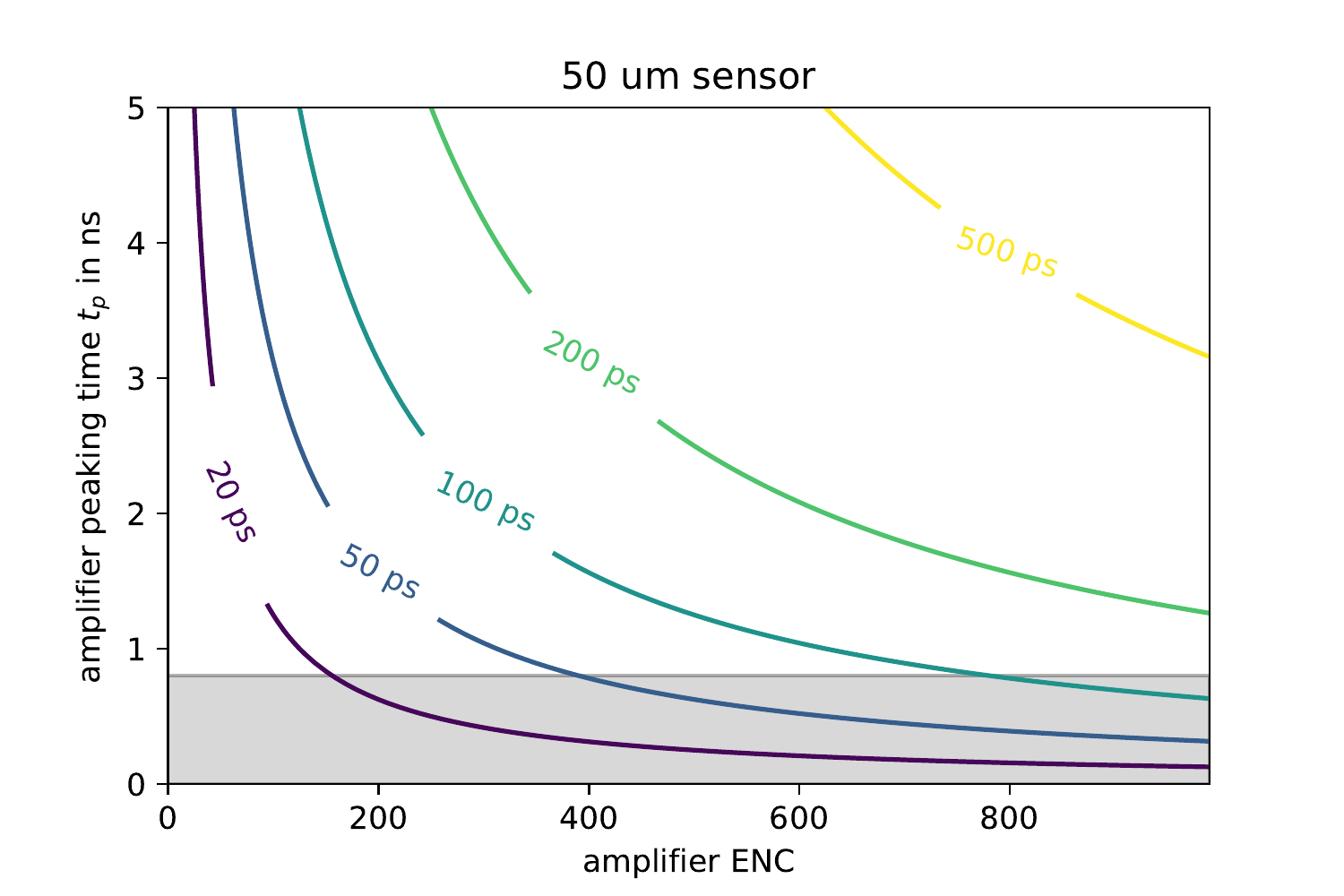}  
  \label{fig:preamp_enc_50}
\end{subfigure}
\caption{Preamplifier requirements on effective noise charge (ENC) and peaking time to achieve a given time resolution with a silicon sensor without internal amplification. The shaded area indicates the signal collection time for a sensor of the given thickness. Calculation based on \cite{riegler}.}
\label{fig:preamp_enc}
\end{figure}

The primary goal of such a dedicated project would be to develop a prototype pixel structure with an integrated preamplifier close to the thermodynamic limit in noise per power consumption. \autoref{fig:preamp_enc} shows the required effective noise charge of an in-pixel amplifier for different target time resolutions and silicon sensor thicknesses.

While it is unlikely for any realistic fast MAPS sensor to reach the timing performance of (AC-)LGADs, it should be possible to get close to it within a factor of two. While LGADs or other future sensors might remain the technology choice for applications demanding the highest possible MIP timing resolution, the prospect of achieving almost comparable resolutions with significantly reduced cost is extremely appealing. Opening up the fast timing frontier for MAPS also enables future MAPS tracker projects to include as much time resolution into their designs as needed for the application. Establishing fast MAPS is thus of prime importance not only for HEP detectors, but for technologically related instrumentation as a whole. 

Fast MAPS have the potential to fulfill all needed requirements for a STOPGAP system or a timing layers in an upgraded inner tracking system. Since the technical requirements, especially for STOPGAP, are fairly tame apart from the MIP time resolution, it poses an ideal prototype project to establish fast MAPS for HEP instrumentation. 

\section{Resource estimates and timeline}
The resource estimate for the whole project is divided into two parts: the development of the MAPS technology, and the actual construction of the detector modules. The given numbers come with great uncertainties and should be considered as rough guidelines.

\subsection{R\&D for fast MAPS}
Making significant steps towards very fast CMOS sensors with timing resolution around \SI{50}{ps} is the most important fundamental building block of this proposal. In that regard the goal of building a STOPGAP system is a vehicle to motivate the extensive R\&D needed towards such fast CMOS sensors. On the way, the current practical limitations of this technology choice will be discovered and, eventually, overcome.

The current efforts towards achieving the best time resolution with CMOS sensors use fabrication processes of \SI{130}{\nm} and \SI{150}{\nm} structure size. Since the noise figure is correlated to the process feature size, an immediate improvement in time resolution (or alternatively the power draw at a fixed time resolution) can be achieved moving the development to the next smaller node of \SI{65}{\nm}. To this end, a suitable \SI{65}{\nm} process with sufficiently high surface resistivity to support high voltage biasing needs to be identified and qualified first. An general drawback of \SI{65}{\nm} processes respect to \SI{130}{\nm} or \SI{150}{\nm} ones are the higher costs per submission and the lower availability of multi project wafer shuttle projects, increasing the costs of such developments.
At least two engineering submissions and extensive laboratory test bench and testbeam campaigns will be necessary to develop and validate a sensor layout that achieves the necessary time resolutions at an acceptable power budget. 

The second step is to integrate TDC electronics and the readout logic onto the same chip. Recent developments in \SI{65}{\nm} processes have demonstrated TDCs with bin widths \SI{<10}{ps} \cite{etroc_tdc, picotdc}. We thus do not expect fundamental issues in designing an on-chip TDC that reaches the resolution range of \SIrange{20}{30}{ps} needed to read out our sensors.

The intermediate goal of this proposed R\&D program is the construction of a small sized prototype detector with active area of a few \si{\cm\squared} in order to validate the sensor technology in a realistic environment. This prototype will be installed inside Belle\,II and operated synchronously with the other detectors while recording physics collisions. The readout of the prototype does not necessarily need to be constructed using a final chip with fully integrated electronics, but could e.g. be read out using a picoTDC \cite{picotdc} integrated into the prototype module. Demonstrating the time-of-flight resolution for single MIPs in the environment of an active HEP experiment will not only validate the sensor technology, but also uncover the practical issues with operating a sensor in realistic conditions compared to test beam or bench tests. The ultimate capstone of the program would be a working prototype of a fully integrated monolithic detector module with integrated sparsified readout that demonstrates the required timing and hit rate capabilities at least in a test beam environment. 

We believe that such a program is not feasible without significant funding for at least three to five years. Out of that time, we coarsely estimate around one to two years will be needed to finish the sensor design, at least one year is needed for the integrated timing frontend and readout logic, and an additional year will be needed for integration onto one common mixed signal chip. Depending on available person power and expertise, parts of this design and test effort can run in parallel. To acquire the necessary momentum, the strong support of an existing local work group with extensive expertise in CMOS sensor design and a profound interest in pushing the possibilities of fast MAPS to its limits is required at the very least for the first one to two years.

\subsection{STOPGAP construction}
Ultimately, 16 STOPGAP modules with around \SIrange{1}{2}{\m\squared} of total active sensor area will have to be built and installed. Apart from the significant funding requirements for the sensor production itself, this will need detailed concepts for the mechanics, cooling, service routing and installation procedure for these modules. None of these tasks have even been considered so far. First steps towards a STOPGAP module design in the form of finite element simulations to estimate the required stiffness and eventual mechanical mock-ups could in principle start in parallel with the sensor R\&D program outlined above.

\section{Summary and Outlook}
Around \SI{6}{\percent} of the charged tracks in the nominal coverage of the TOP particle identification system in the Belle\,II barrel region pass near or through the gaps between the TOP quartz radiator bars and thus do not yield usable PID information. An additional \SI{10}{\percent} of tracks suffer from degraded PID performance from passing too close to the quartz edges. The STOPGAP project aims to fill these gaps in coverage with fast-timing silicon sensors to measure the time of flight of such particles to improve the Belle\,II particle identification coverage. Examining the possible high precision timing technologies, CMOS timing detectors appear to be most promising for the requirements of a Belle II upgrade. We propose to start a significant CMOS timing sensor R\&D program with a series of prototypes and beam tests. The intermediate goal would be to integrate a small sensor protoype into Belle\,II and to operate it in physics collisions.

\appendix
\section{Appendix}

\begin{figure}[htbp]
	\centering
    \includegraphics[width=0.7\textwidth]{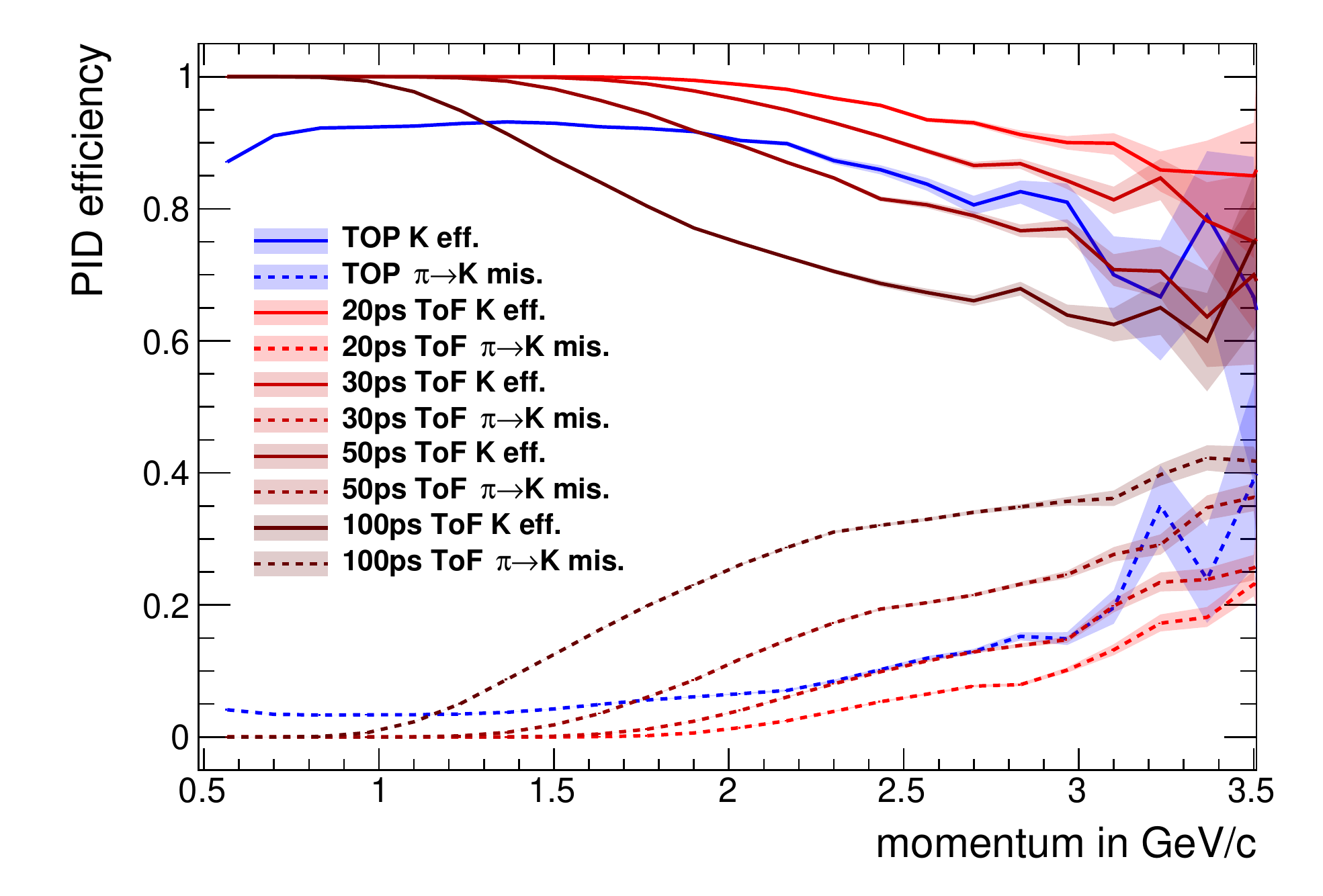}
	
	\caption[]{Kaon selection and mis-identification probabilities at $LL_{K}-LL{\pi}>0$ of a simulated STOPGAP PID system for charged kaons as a function of particle momentum, shown for simulated sensor resolutions in the range of \SIrange{20}{100}{ps} MIP timing. Solid curves are the kaon selection efficiencies, while the dashed curves show the mis-identification probability for $\pi\rightarrow K$. Shaded areas indicate the statistical uncertainties of the data points.}
	\label{fig:tof_perf_kaon}
\end{figure}

\begin{figure}[htbp]
	\centering
    \includegraphics[width=0.7\textwidth]{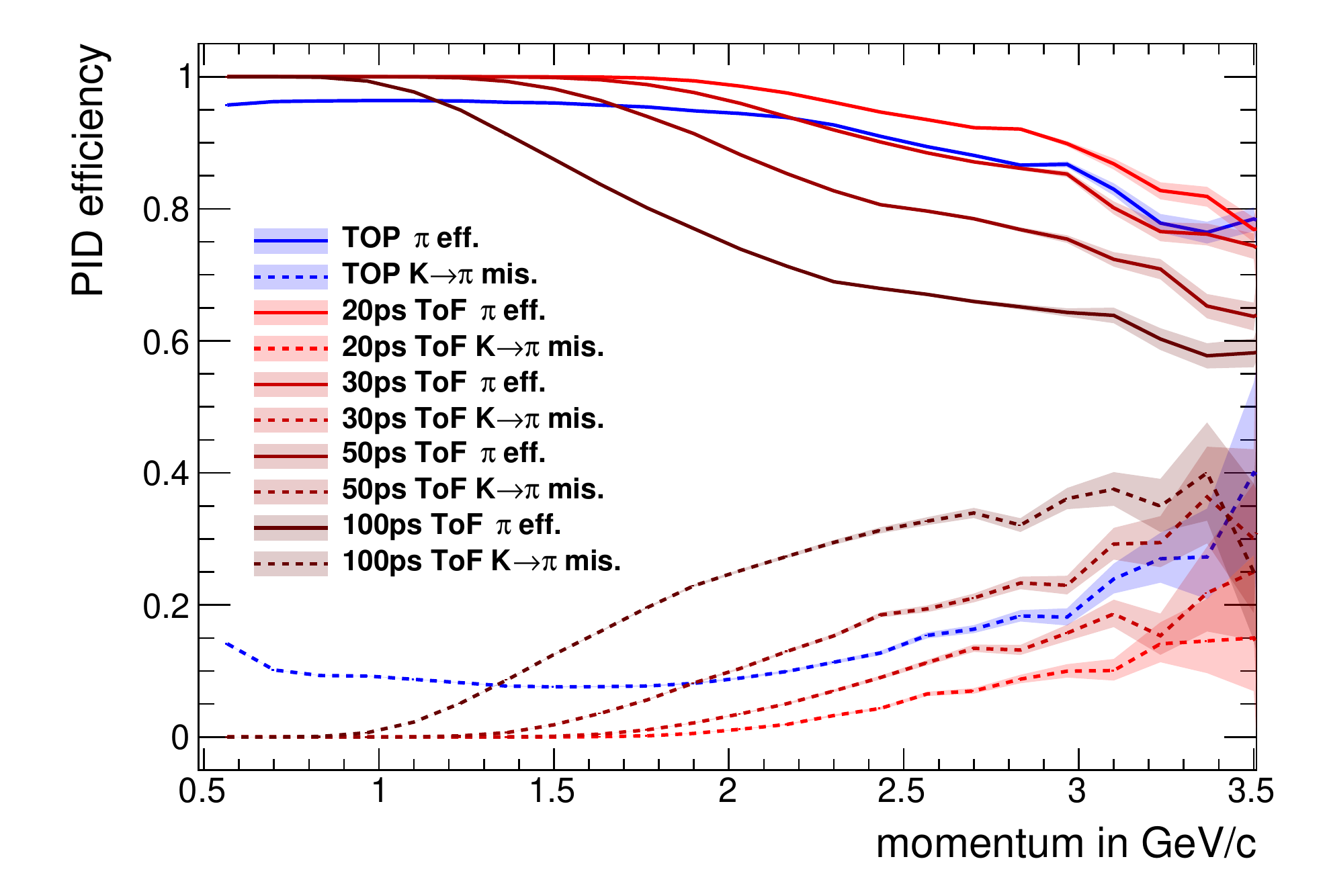}
	
	\caption[]{Pion selection and mis-identification probabilities at $LL_{K}-LL{\pi}>0$ of a simulated STOPGAP PID system for charged kaons as a function of particle momentum, shown for simulated sensor resolutions in the range of \SIrange{20}{100}{ps} MIP timing. Solid curves are the kaon selection efficiencies, while the dashed curves show the mis-identification probability for $\pi\rightarrow K$. Shaded areas indicate the statistical uncertainties of the data points.}
	\label{fig:tof_perf_pion}
\end{figure}

\end{document}